\documentclass{jetpl}
\twocolumn

\lat

\title{Atomic structure of Ge quantum dots on the Si(001) surface}

\rtitle{Atomic structure of Ge quantum dots on the Si(001) surface}

\sodtitle{Atomic structure of Ge quantum dots on the Si(001) surface}

\author{L.\,V.\,Arapkina,
 V.\,A.\,Yuryev\thanks{http://www.gpi.ru/eng/staff\_s.php?eng=1\&id=125, e-mail: vyuryev@kapella.gpi.ru}\/}

\rauthor{Arapkina L.\,V., Yuryev V.\,A.}

\sodauthor{Arapkina, Yuryev}

\address{A.\,M.\,Prokhorov General Physics Institute of the Russian Academy of Sciences,\\ 38 Vavilov Street, Moscow, 119991, Russia}

\dates{*}{*}

\abstract{
{\em In situ} morphological investigation of the \{105\} faceted Ge islands on the Si(001) surface (hut clusters) have been carried out using an ultra high vacuum  instrument integrating a high resolution scanning tunnelling microscope and a molecular beam epitaxy vessel. Both species of hut clusters---pyramids and wedges---were found to have the same structure of the \{105\} facets which was visualized. Structures of vertexes of the pyramidal clusters and ridges of the wedge-shaped clusters were revealed as well and found to be different. This allowed us to propose a crystallographic model of the \{105\} facets as well as models of the atomic structure of both species of the hut clusters. An inference is made that transitions between the cluster shapes are impossible.

}


\begin{document}

\maketitle

\begin{figure*}
\includegraphics[scale=1.5]{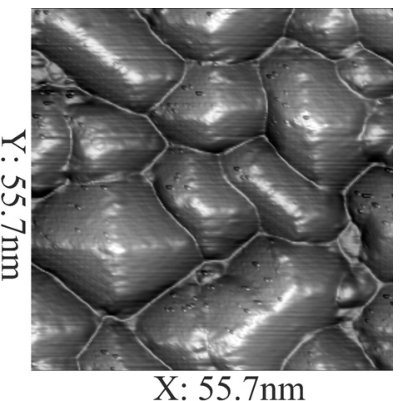}(a)
\includegraphics[scale=1.5]{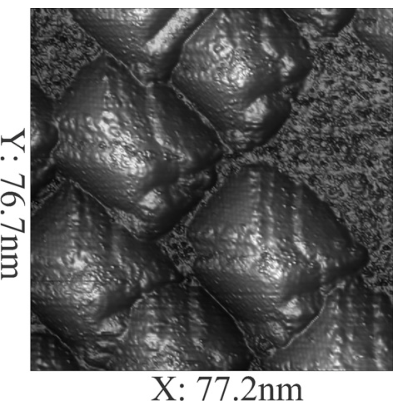}(b)
\caption{\label{fig:arrays}Fig.1 
STM images of arrays of Ge hut clusters grown on the Si(001) surface,  $h_{\rm Ge}=10$~\r{A},  $T_{\rm gr}$ is (a) $360^{\circ}$C ($U_{\rm s}=+2.5$~V, $I_{\rm t}=80$~pA) and (b) 
 $530^{\circ}$C ($U_{\rm s}=+2.1$~V, $I_{\rm t}=80$~pA).}
\end{figure*}

Dense arrays of the self-assembled Ge quantum dots (QD)  deposited on the Si(001) surface at moderate temperatures and compatible with the Si technology (Fig.~\ref{fig:arrays}) stay an attractive subject of investigations for a number of years (see e.g. Refs.~\cite{Mo,Chem_Rev,Kastner,Fujikawa,Pyramid_to_dome}) due to their promising potential applications in microelectronics and primarily in microphotoelectronics \cite{Wang-Cha,Wang-properties}. Significant technological achievements of the recent years enabled the controllable formation of the QD arrays with the desired cluster densities \cite{Smagina,classification}. However, the uniformity of cluster types in the arrays (or array defectiveness) and the dispersion of cluster  sizes are still  issues of the day. 

It was shown recently that the  \{105\} faceted clusters usually referred to as hut clusters are subdivided into two main species---pyramids and wedges. Their densities are equal at the initial stage of the array formation. Then, as the Ge coverage is increased, the later become dominating in the arrays whereas the former exponentially rapidly disappear   
\cite{classification}. (The authors of Ref.~\cite{Kastner} observed the analogous process---the growth of Ge coverage resulted in the increase of density of elongated hut clusters.) 
An additional peculiarity is the difference of heights of the pyramidal and wedge-shaped clusters. The formed pyramids are always higher than the  wedges and the later are limited in their heights \cite{classification}. The limit height depends on the growth temperature.
The dispersion of the cluster sizes in the arrays was found to be governed by the lengths of the wedge-like clusters which appeared to be distributed rather uniformly over a wide interval of values of the cluster base aspect ratio~\cite{classification}. 
In order to control the parameters of the  arrays a deep knowledge of the cluster morphology is important.

In this article, we investigate the structure of both main species of the hut clusters on the atomic level. 
It can be shown based on the bulk Ge lattice that the \{105\} facet is composed by monoatomic steps and (001) terraces \cite{Kastner}. The step height is $\sim 1.4$~\r{A}, the terrace width is $\sim 7$~\r{A}. The first model of the hut cluster faces was proposed in Ref.~\cite{Mo}. Later another model appeared in Ref.~\cite{Fujikawa}. Both models are founded on the assumption that a reconstruction formed by the Ge ad-dimers is present on the free surfaces of the (001) terraces. The main difference between the models is in the determination of this reconstruction. Now a new model of the  hut cluster faces is brought forward on the basis of our {\it in situ} investigations of the \{105\} facet carried out by the high resolution scanning tunnelling microscopy (STM). In addition, crystallographic models of pyramidal and wedge-shaped clusters are proposed based on the facet structure and a fine structure of the vertices of the pyramids  and the ridges of the wedges.

\begin{figure*}
\includegraphics[scale=1.5]{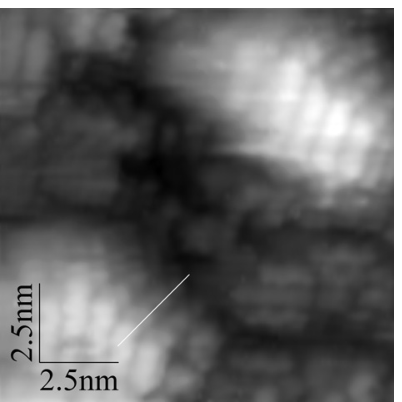}(a) 
\includegraphics[scale=1.5]{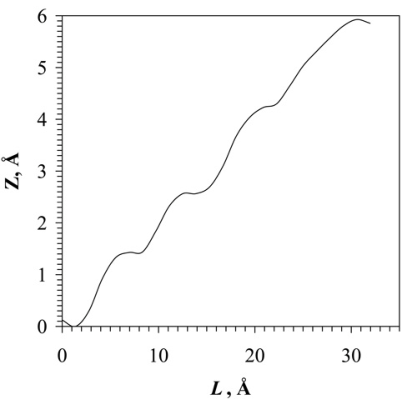}(b) 
\caption{\label{fig:profile} Fig.2 
STM image  (a) of  Ge hut clusters ($h_{\rm Ge}=6$~\r{A}, $T_{\rm gr}=360^{\circ}$C,
 $U_{\rm s}=+1.8$~V, $I_{\rm t}=80$~pA),  
 blocks of the Ge wetting layer $(M\times N)$ structure with the $p(2\times 2)$ and $c(4\times 2)$ reconstructions \cite{Chem_Rev} are seen in the lower right quarter of the field. A  profile of the cluster facet (c) taken along the white line shown in the image 
(a), 
the monoatomic steps and (001) terraces are clearly seen.}
\end{figure*}


The experiments were carried out using an ultra high vacuum instrument consisting of the molecular beam epitaxy (MBE) chamber coupled with STM which enables the sample study at any stage of processing serially investigating the structure and giving additional treatments to the specimen. 
 The surface of the silicon substrates was completely purified of the oxide film as a result of short annealing at the temperature of $\sim 925^\circ$C \cite{our_Si(001)_en}. Germanium was deposited directly on the purified silicon (001) surface from the source with the electron beam evaporation.
The rate of Ge deposition was  $\sim 0.15$~\r{A}/s, the Ge coverage  ($h_{\rm Ge}$) (or more accurately the thickness of the Ge film measured by the graduated in advance film thickness monitor with the quartz sensor installed in the MBE chamber) was varied from 6 to 14~\r{A} for different samples.  The substrate temperature $T_{\rm gr}$ was 360 or $530^\circ$C during the process. The rate of the sample cooling down to the room temperature was  $\sim 0.4^\circ$C/s after the deposition. Images were scanned at room temperature in the constant tunnelling current ($I_{\rm t}$) mode. The STM tip was zero-biased while a sample was positively or negatively biased ($U_{\rm s}$). The details of the sample preparation as well as the experimental techniques can be found in Refs.~\cite{classification} and \cite{our_Si(001)_en}. The WSxM software \cite{WSxM} was used for processing of the STM images.


\begin{figure*}
\includegraphics[scale=1.5]{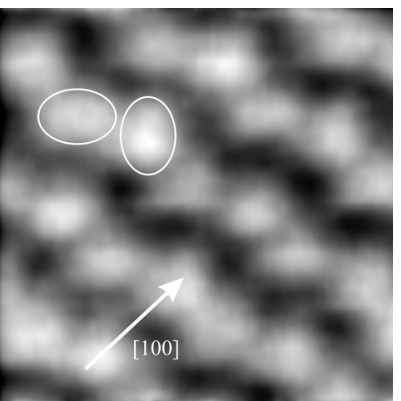}(a)
\includegraphics[scale=1.5]{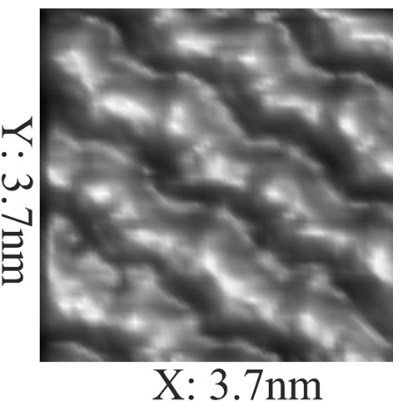}(b)
\caption{\label{fig:facet} Fig.3 
2-D (a) and   3-D (b) STM images  of the same area on Ge hut cluster facet ($h_{\rm Ge}=10$~\r{A}, $T_{\rm gr}=360^{\circ}$C,
$U_{\rm s}=+2.1$~V, $I_{\rm t}=80$~pA). The sides of the cluster base lie along the [100] direction, structural units arising on the free surfaces of the (001) terraces are marked out. }
\end{figure*}

\begin{figure*}
\includegraphics[scale=1.4]{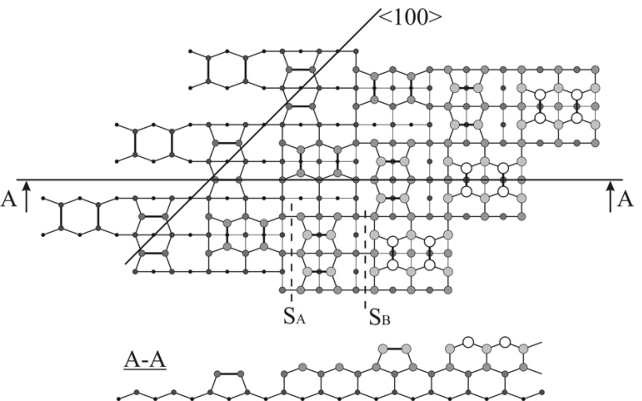}(a)
\includegraphics[scale=1.4]{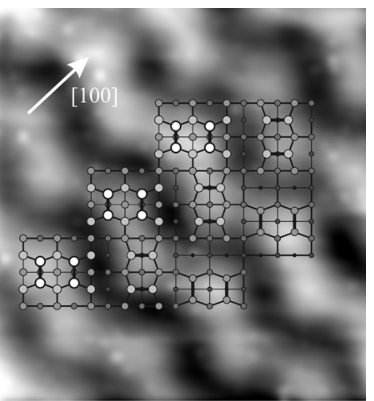}\,(b)
\caption{\label{fig:face_structure}Fig.4
 A structural model of the $\{105\}$ facet of hut clusters (a),  S$_{\rm A}$ and  S$_{\rm B}$ are commonly adopted designations of the monoatomic steps \cite{Chadi}, atoms situated on higher terraces are shown by larger circles.
(b) The schematic of the  facet  superimposed on its STM image  ($4.3\times 4.4$~nm, $U_{\rm s}=+3.0$~V, $I_{\rm t}=100$~pA), the [100] direction coincides with that of the cluster base side.}
\end{figure*}


\begin{figure*}
\includegraphics[scale=1.3]{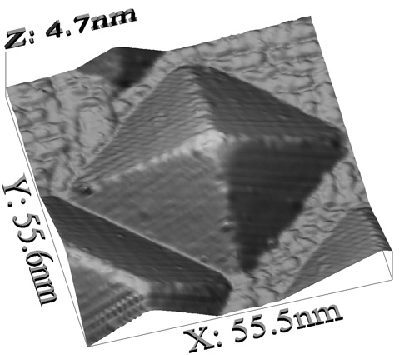}(a)
\includegraphics[scale=1.3]{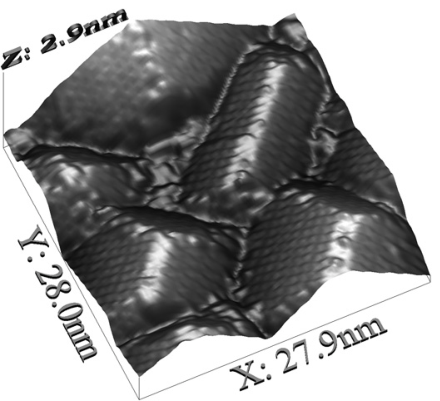}(b)
\includegraphics[scale=1.3]{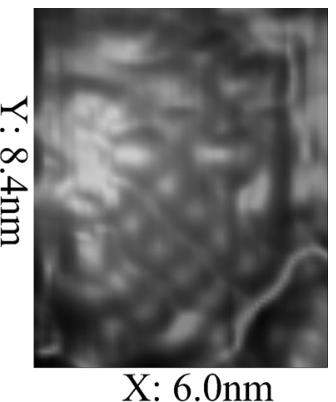}(c)
\includegraphics[scale=1.3]{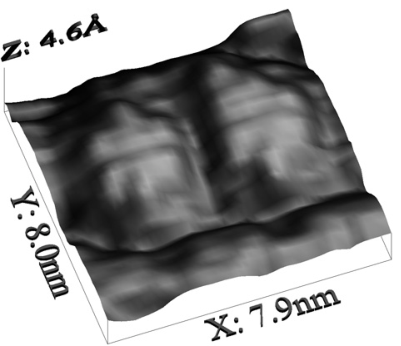}(d)
\includegraphics[scale=1.3]{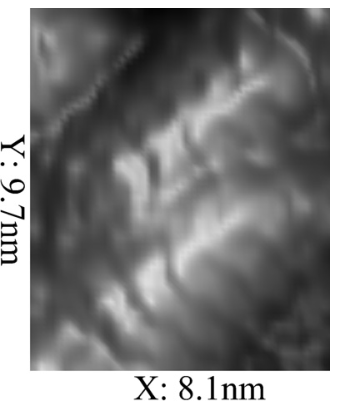}(e)
\includegraphics[scale=1.3]{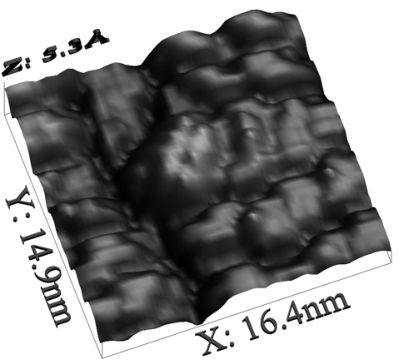}(f)
\caption{\label{fig:top_view}Fig.5
STM topographs of (a) pyramidal ($T_{\rm gr}$ = 530$^\circ$C, $h_{\rm Ge} = 11$~\AA)
and (b) wedge-shaped ($T_{\rm gr}$ = 360$^\circ$C, $h_{\rm Ge} = 10$~\AA)
clusters, (c)
the   vertex, face  and (d) a nucleus [\,the left of two features, 1\,ML over WL\,]  of a pyramid ($T_{\rm gr} = 360^\circ$C, $h_{\rm Ge} = 6$~\r{A}), (e) the
ridges and  long facets  of two closely neighbouring wedges  ($T_{\rm gr} = 360^\circ$C, $h_{\rm Ge} = 8$~\r{A}) and (f) a small wedge [\,in the center of the field of view, 2\,ML over WL\,]  ($T_{\rm gr} = 360^\circ$C, $h_{\rm Ge} = 6$~\r{A}).}
\end{figure*}

\begin{figure*}
\includegraphics[scale=1.95]{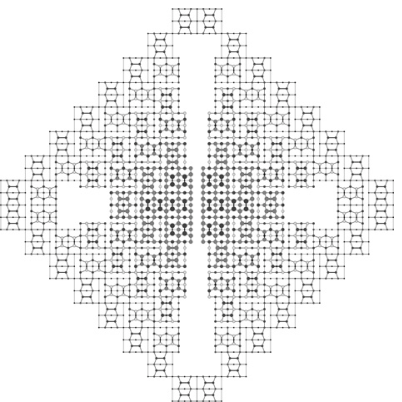}(a)
\includegraphics[scale=1.95]{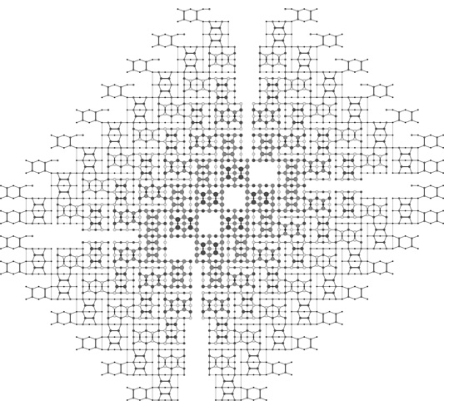}(b)
\caption{\label{fig:schematic}Fig.6
Schematic drawings of atomic structures of Ge (a) pyramidal and (b) wedge-shaped   hut clusters composed of 6 monoatomic steps on the wetting layer.
}
\end{figure*}

The obtained experimental results are as follows.

Fig.~\ref{fig:arrays} demonstrates STM images of the Ge QD arrays formed at 360 and 530$^\circ$C and the same value of $h_{\rm Ge} = 10$~\r{A}. Ge clusters grown at these temperatures and the mentioned value of  $h_{\rm Ge}$ are seen to have different structure of faces---the facets of the clusters grown at 360$^\circ$C are completed whereas the faces of the clusters formed at 530$^\circ$C appeared to be uncompleted and contain additional steps of the solidifying interfaces (Fig.~\ref{fig:arrays}(b)). This means that Ge clusters reached some limit height determined by the growth temperature begin to non-uniformly ad the  migrating Ge atoms  to the faces. As a result new incomplete planes are formed on the facets. No small clusters or nuclei were found on the wetting layer (WL) in the arrays grown at  530$^\circ$C. Simultaneous growth and nucleation of several phases of Ge clusters is typical for the arrays growing at lower temperatures (e.g. 360$^\circ$C) \cite{classification}. So at 530$^\circ$C the process of the QD growth dominates over the process of nucleation of new clusters after the  act of the initial cluster nucleation whereas at 360$^\circ$C these processes compete during the array growth. The details of classification of Ge hut clusters and the properties of QD arrays grown at moderate temperatures can be found in Ref.~\cite{classification}.

Fig.~\ref{fig:profile} demonstrates an STM image of a small Ge QD, the wetting layer in the nearest vicinity of its base (a) and a profile of its facet (b). The WL consists of $(M\times N)$ blocks with the $p(2\times 2)$ and $c(4\times 2)$ reconstruction (Fig.~\ref{fig:profile}(a) and Ref.~\cite{Chem_Rev}). The QD is seen  to be formed by the monoatomic steps (inclined regions of the curve) and the (001) terraces (horizontal regions of the curve). Heights of the steps are $\sim 1.4$~\r{A} and total widths of the terraces are  $\sim 7$~\r{A} that confirms the assumption made in Ref.~\cite{Kastner} regarding the sizes of the steps and terraces.  
The line connecting the maximum and the minimum of the profile has a slope of $\sim$~11.3$^\circ$.
The slopes of the inclined regions of the profile are much greater.

The STM investigations of faces of the clusters grown at 360 and 530$^\circ$C evidenced that their fine structure depends on neither the growth temperature nor the hut cluster species. A typical STM image of the QD facet is presented in Fig.~\ref{fig:facet}. Characteristic distances on the facets are as follows: $\sim 10.5$~\r{A} in the \textless 100\textgreater~directions (along the corresponding side of the base) and  $\sim 14$~\r{A} in the normal (\textless 051\textgreater) directions.
The facets are composed by structural units which are outlined by ellipses in Fig.~\ref{fig:facet}(a) and can be arranged along either [110] or [1${\overline 1}$0] direction on the (001) plane. The spatial arrangement of the units  is changed to perpendicular on the adjacent terraces and remains unchanged within the terrace. Following the authors of Refs.~\cite{Mo} and \cite{Fujikawa}, we suppose this structural unit to be a pair of dimers composed by Ge atoms on the terrace surface. The QD itself may be considered as a set of successive monoatomic steps and (001) terraces which form the \{105\} faces \cite{Mo,Chem_Rev,Kastner,Fujikawa}.
We observe pairs of dimers which are the ``reminders'' of dimer rows covering the free ``outcrops'' of the (001) planes. As the dimer rows change their direction to the perpendicular on the adjacent terraces, the observed structural units turn at 90$^\circ$ as well.

A model of facets of the Ge hut clusters  may be built on the basis of the above experimental observations and the previous data of Refs.~\cite{Mo} and \cite{Fujikawa}. The model is based on the following assumptions: (i) the hut clusters are faceted by the \{105\} planes, (ii)  the hut cluster faces consist of the monoatomic steps and (001) terraces, and (iii) a structure is formed on the free ``outcrops'' of the (001) terraces which consists of pairs of ad-dimers and is similar to the structures arising on the pure Si(001) or Ge(001) surfaces \cite{Chadi}. A schematic of this model is shown in Fig.~\ref{fig:face_structure}(a). The sides of the cluster base are alined with the \textless 100\textgreater~directions. The terraces are bounded by the S$_{\rm A} $ and  nonrebonded S$_{\rm B} $ monoatomic steps (the terminology by Chadi \cite{Chadi}). Fig.~\ref{fig:face_structure}(b) demonstrates the drawing of the structure superposed with its STM image.

As mentioned above there are two main species of Ge hut clusters---pyramidal and wedge-shaped ones \cite{classification}. Images of the clusters of both species are shown in Figs.~\ref{fig:top_view}(a,\,b). It is seen that the structures of the top vertex of the pyramid  and the top edge of the wedge  (the ridge) are different. To build models of atomic structure of Ge hut clusters the knowledge of architecture of  tops of each species is required. Figs.~\ref{fig:top_view}(c,\,d) demonstrate high resolution topographs  of the vertex and  a nucleus (1~ML over WL) of the pyramid-like  cluster and Figs.~\ref{fig:top_view}(e,\,f)  show  the topographs of the ridge and a very small (only 2~ML over WL) wedge-like QD.
These images allowed us to find out the required structures of tops which are shown as the highest atomic layers in the structural schematics of each species of the clusters  demonstrated in Fig.~\ref{fig:schematic}. 
We used the following procedure to build the models of the clusters: (i) we drew each of four facets in accordance with the schematic shown in Fig.~\ref{fig:face_structure}(a)  (ii) then we combined the faces according to the results of the STM study of tops. For the simplicity and clarity of the drawings (Fig.~\ref{fig:schematic}) the structure of edges is not shown and only the highest layer of atoms is shown on the vertex and the ridge.

It should be remarked in the conclusion that according to our STM data pyramides and wedges have different structures of tops. It is seen from Figs.~\ref{fig:top_view} and \ref{fig:schematic} that one cannot transform the structure of clusters of one species to the structure of the other. The structure of the pyramid cannot be obtained from the structure of the wedge simply combining four equal faces as the resultant model would have a wrong structure of the vertex. Hence, the conditions and processes of nucleation and growth  should be expected to be different for pyramidal and wedge-like Ge hut clusters.

In summary, we investigated the structure of the \{105\} faceted Ge quantum dots (hut clusters) grown on the Si(001) surface at the temperatures of 360 and 530$^\circ$C. Despite the difference of the formation processes at these temperatures both species of hut clusters (pyramidal and wedges-like ones) were found to always have the same structure of the \{105\} facets which was visualized. Structures of the vertexes of the pyramidal clusters and the ridges of the wedge-shaped clusters were revealed as well. This allowed us to bring forward a crystallographic model of the \{105\} facets as well as the models of the atomic structure of both species of Ge hut clusters.


\vfill\eject
\newpage

Fig.1. STM images of arrays of Ge hut clusters grown on the Si(001) surface,  $h_{\rm Ge}=10$~\r{A},  $T_{\rm gr}$ is (a) $360^{\circ}$C ($U_{\rm s}=+2.5$~V, $I_{\rm t}=80$~pA) and (b) 
 $530^{\circ}$C ($U_{\rm s}=+2.1$~V, $I_{\rm t}=80$~pA).

Fig.2. STM image  (a) of  Ge hut clusters ($h_{\rm Ge}=6$~\r{A}, $T_{\rm gr}=360^{\circ}$C,
 $U_{\rm s}=+1.8$~V, $I_{\rm t}=80$~pA),  
 blocks of the Ge wetting layer $(M\times N)$ structure with the $p(2\times 2)$ and $c(4\times 2)$ reconstructions \cite{Chem_Rev} are seen in the lower right quarter of the field. A  profile of the cluster facet (c) taken along the white line shown in the image 
(a), 
the monoatomic steps and (001) terraces are clearly seen.

Fig.3. 2-D (a) and   3-D (b) STM images  of the same area on Ge hut cluster facet ($h_{\rm Ge}=10$~\r{A}, $T_{\rm gr}=360^{\circ}$C,
$U_{\rm s}=+2.1$~V, $I_{\rm t}=80$~pA). The sides of the cluster base lie along the [100] direction, structural units arising on the free surfaces of the (001) terraces are marked out. 

Fig.4.  A structural model of the $\{105\}$ facet of hut clusters (a),  S$_{\rm A}$ and  S$_{\rm B}$ are commonly adopted designations of the monoatomic steps \cite{Chadi}, atoms situated on higher terraces are shown by larger circles.
(b) The schematic of the  facet  superimposed on its STM image  ($4.3\times 4.4$~nm, $U_{\rm s}=+3.0$~V, $I_{\rm t}=100$~pA), the [100] direction coincides with that of the cluster base side.

Fig.5. STM topographs of (a) pyramidal ($T_{\rm gr}$ = 530$^\circ$C, $h_{\rm Ge} = 11$~\AA)
and (b) wedge-shaped ($T_{\rm gr}$ = 360$^\circ$C, $h_{\rm Ge} = 10$~\AA)
clusters, (c)
the   vertex, face  and (d) a nucleus [\,the left of two features, 1\,ML over WL\,]  of a pyramid ($T_{\rm gr} = 360^\circ$C, $h_{\rm Ge} = 6$~\r{A}), (e) the
ridges and  long facets  of two closely neighbouring wedges  ($T_{\rm gr} = 360^\circ$C, $h_{\rm Ge} = 8$~\r{A}) and (f) a small wedge [\,in the center of the field of view, 2\,ML over WL\,]  ($T_{\rm gr} = 360^\circ$C, $h_{\rm Ge} = 6$~\r{A}).

Fig.6. Schematic drawings of atomic structures of Ge (a) pyramidal and (b) wedge-shaped   hut clusters composed of 6 monoatomic steps on the wetting layer.

\end{document}